# Temperature dependence of the structural parameters of the non-oxide perovskite superconductor $MgCNi_3$

Q. Huang[1], T. He[2], K. A. Regan[2], N. Rogado[2], M. Hayward[2], M.K. Haas[2], K. Inumaru[2], and R.J. Cava[2]

[1]*Department of Materials and Nuclear Engineering, University of Maryland, College Park, MD 20742, and NIST Center for Neutron Research, Gaithersburg, MD 20899,*

[2]*Department of Chemistry and Princeton Materials Institute, Princeton University, Princeton NJ 08544*

## Abstract

We report the structural parameters of superconducting $MgC_xNi_3$ ($x$=0.96, $T_C$=7.3 K) as a function of temperature, from 2 K to 295 K, determined by neutron powder diffraction profile refinement. The compound has the perovskite structure over the whole temperature range, with symmetry $Pm\bar{3}m$ and $a$=3.81221(5) Å at 295 K: no structural or long range magnetic ordering transitions were observed. The lattice parameter $a$ and the Debye-Waller factors for the individual atoms decrease smoothly with decreasing temperature. There are no unusual changes of the structural parameters near $T_C$.

## Introduction

The recent discovery of superconductivity at 39 K in $MgB_2$ [1] has suggested that intermetallic compounds should be reconsidered as a potential source of high temperature superconducting materials. Subsequent to that discovery, the observation of superconductivity at temperatures near 8 K was made in $MgCNi_3$ [2]. The superconducting transition temperature is too low to be of practical value, but the high proportion of Ni in the compound suggests that magnetic interactions may be important to the existence of the superconductivity. $MgCNi_3$, with the cubic perovskite structure, can be considered as the three-dimensional analog of the layered nickel borocarbides, typified by $LuNi_2B_2C$, which has a $T_c$ near 16 K [3]. Here we report the temperature dependence of the structural parameters of $MgCNi_3$ between ambient temperature and 2 K, to determine whether it undergoes any of the symmetry-lowering transitions commonly seen in oxide perovskites on cooling, and whether there are any structural anomalies at the superconducting transition temperature. The results show, with very high precision, that the thermal behavior of $MgCNi_3$ on cooling is of a conventional nature, with no structural transitions observed, and no detectable anomalies at $T_c$.

## Experiments and results

For the neutron diffraction study, a 3 gram powder sample of nominal composition $MgC_{1.25}Ni_3$ was prepared, as described elsewhere [2]. The superconducting transition was characterized magnetically, and found to be at 7.3 K [2]. The neutron powder diffraction intensity data were collected using the BT-1 high-resolution powder diffractometer located at the reactor of the NIST Center for Neutron Research, employing a Cu311 monochromator to produce a monochromatic neutron beam of wavelength 1.5401 Å. Collimators with horizontal divergences of 15', 20', and 7' arc were used before and after the monochromator, and after the sample, respectively. The intensities were measured in steps of 0.05° in the 2θ range 3°-168°. Data were collected at different temperatures from 2K to 300 K. Structural parameters were refined by using the program GSAS [4]. The neutron scattering amplitudes used in the refinements were 0.538, 0.665, and 1.03 ($\times 10^{-12}$cm) for Mg, C and Ni, respectively. The ambient temperature neutron powder diffraction pattern indicated that $MgC_{0.96}Ni_3$ has $Pm\overline{3}m$ symmetry with $a$=3.812 Å. A few low intensity peaks in the pattern were identified to be due to graphite impurity, consistent with the nominal formula, and were taken into account in the refinements ( ~2 weight % of the sample).

$MgC_{0.96}Ni_3$ was found to have the classical perovskite structure, as shown in fig. 1, with the atomic positions: Mg: 1*a* (0,0,0), C: 1*b* (0.5, 0.5, 0.5) and Ni: 3*c* (0, 0.5, 0.5). The isotropic temperature factors obtained in the refinement are 0.90(3), 0.54(4), and 0.75(1) $Å^2$ for Mg, C, and Ni, respectively at ambient temperature. The final refinements were performed by allowing the occupancy of the C site to vary. This parameter was found to be ~0.960(8) making the exact stoichiometry $MgC_{0.96}Ni_3$. This result is consistent the small amount of unreacted graphite (2 weight %) found in the sample. The refinement agreement factors were $R$=5.14%, $R_{wp}$=6.39% and $\chi^2$= 1.258. The same good fits were obtained for all the data collected at different temperatures. The observed and calculated intensities at 296 K are shown in the Fig. 2.

The inset of Fig. 2 shows, as an example, the observed intensity (crosses) fitted by a Gausian function (solid line) for the reflection (222) at 295 K. The figure indicates that the peak width and shape are close to the instrumental resolution. In order to exclude any possible unusual structural behavior for this compound, the same fitting was performed for all the peaks above 70 degrees 2-theta at different temperatures. Results indicate that no observable particle and/or strain broadening is present in this sample in the measured temperature range, and that no evidence of structural distortion was observed with decreasing temperature.

The refined lattice parameter *a* and Debye-Waller factors decrease smoothly as the temperature decreases. Fig. 3 (upper panel) shows the plots of the lattice parameter *a* as function of temperature over a wide temperature range. The solid curve gives the fitting by a polynomial function $a_T=a_0+\alpha T+\beta T^2$ with $a_0$, $\alpha$ and $\beta$ equal to 3.8066, $3.7985\times10^{-6}$, and $5.3493\times10^{-8}$, respectively. In the equation, $a_T$ is the lattice parameter at temperature T and $a_0$ is its value at T=0 K. The quantities $\alpha$ and $\beta$ are the polynomial coefficients and T is the absolute temperature. Unlike the case for $MgB_2$, the thermal expansion cannot be fit to a model where the behavior is dominated by a single phonon energy [5]. This is as expected because the strongly bonded network of light atoms present in $MgB_2$ (the B honeycomb planes) is not present in $MgCNi_3$. The lower panel of figure 3 shows that careful measurements near $T_c$ (7.3 K) indicate that the lattice parameter shows no anomalies in this temperature range within the experimental uncertainties, which are relatively small.

In the structural refinements, the thermal parameters of Mg and C were refined isotropically, whereas the thermal parameter of Ni was refined in an anisotropically, consistent with the point symmetry for each atom. Fig. 4 shows plots of the Debye-Waller factors vs. temperature. The values of the temperature factors, and their thermal behavior, are similar to those observed in most perovskite compounds. The C atom, at the center of the Ni octahedron, has a smaller temperature factor than the others. The atom (Ni) at the corners of the octahedron has a value of the anisotropic mean square displacement factor ($U_{11}$) significantly smaller in the bonding direction (Ni to C) than in the others ($U_{22}$ and $U_{33}$) in the plane perpendicular to the bonding direction. The lower panel shows that the thermal parameters do not undergo any discontinuous changes in the vicinity of $T_c$ within the precision of our measurements. Finally, the thermal parameters' magnitudes and thermal behavior indicate that the small proportion of vacant positions (4%) on the C sites did not introduce measurable structural disorder.

**Conclusion**

The present study of the temperature dependent structure of the non-oxide perovskite superconductor $MgCNi_3$ has found no magnetic or structural transitions between 2 and 295K. If there is indeed any structural anomaly at $T_c$, it would have to be undetectable to the precision level of these measurements, which are on the order of 0.01%. Further experiments to determine whether magnetic interactions in this simple, high symmetry structure have anything to do with the superconductivity will be of interest, but from a structural viewpoint its behavior is conventional.

**Acknowledgement**

The work at Princeton was supported by the National Science Foundation, and the Department of Energy, Office of Basic Energy Sciences.

**Figure Captions**

Fig 1. The perovskite crystal structure of $MgC_{0.96}Ni_3$. Atom types as marked.

Fig 2. Observed (crosses) and calculated (solid line) neutron powder diffraction intensities for $MgC_{0.96}Ni_3$ at 295 K. Vertical lines indicate the position of all allowed reflections. A difference curve is shown at the bottom. The inset is the (222) reflection at 295 K fitted by a Gausian function.

Fig 3. The cubic lattice parameter *a* of $MgC_{0.96}Ni_3$ vs. temperature. The solid line is least squares fit using a polynomial function (see text).

Fig 4. Debye-Waller factors for each atom in $MgC_{0.96}Ni_3$ , as a function of temperature. The symbols for each parameter used in (b) are the same as those used in (a).

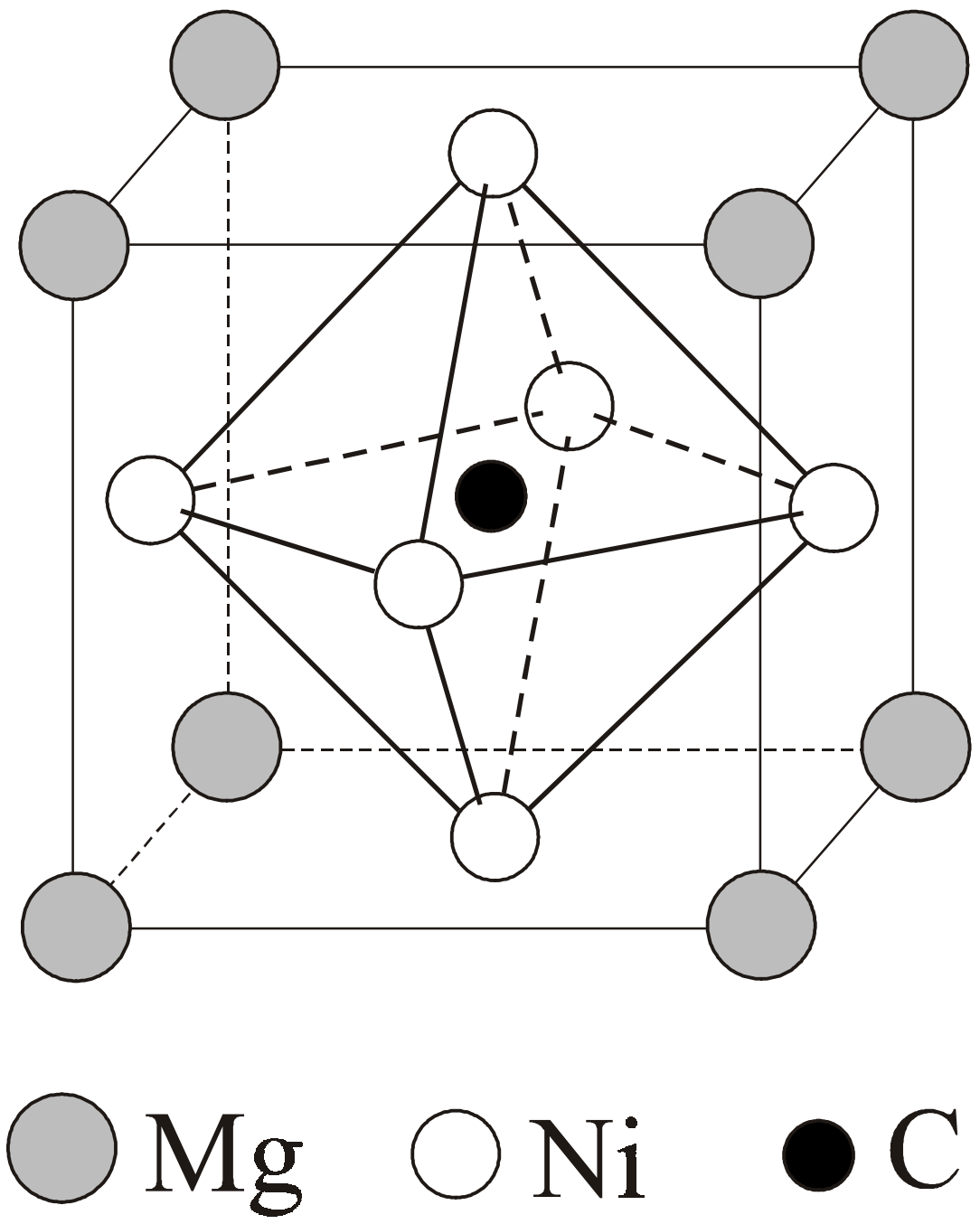

Figure 1 *Huang, et al*

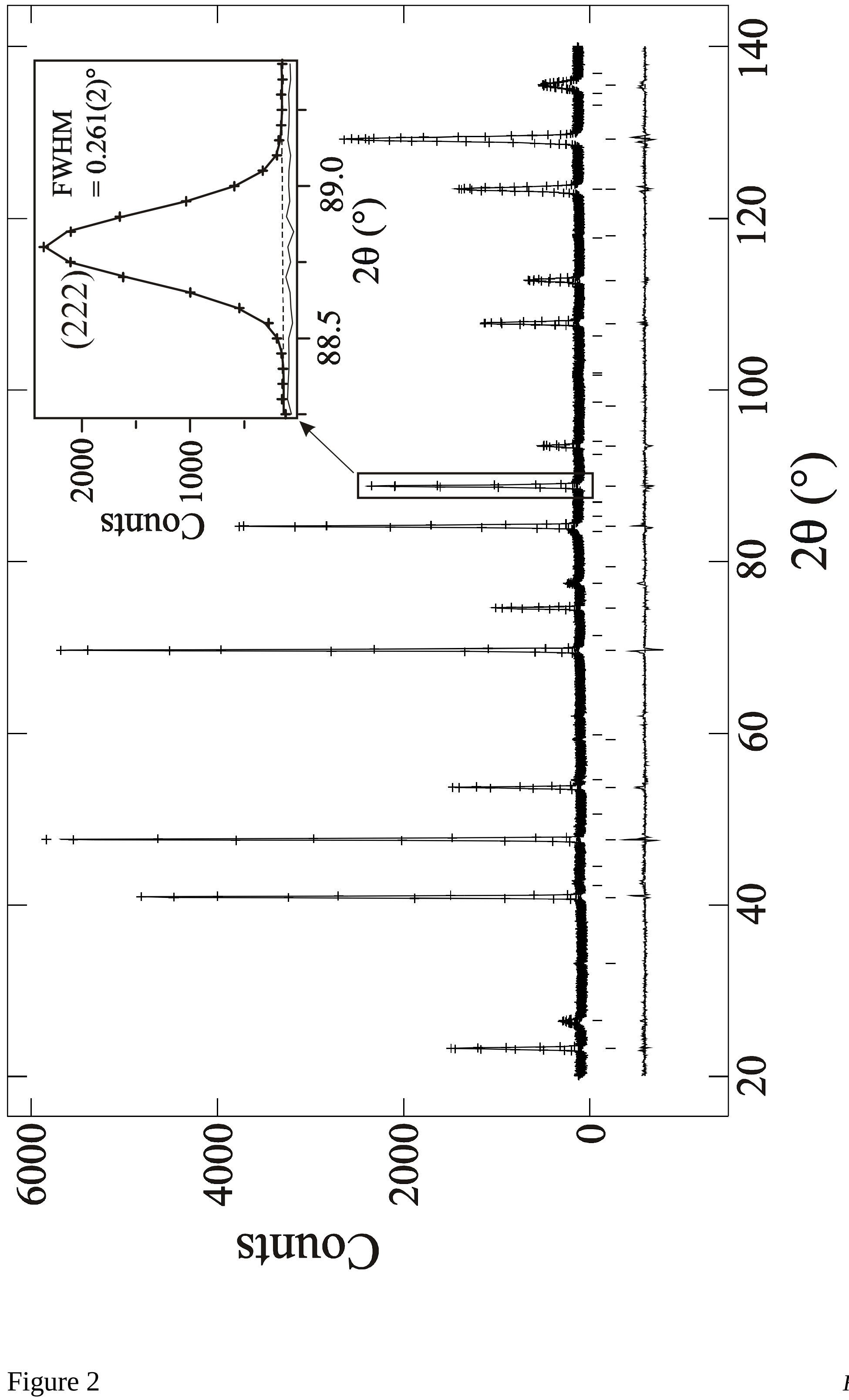

Figure 2 *Huang, et al*

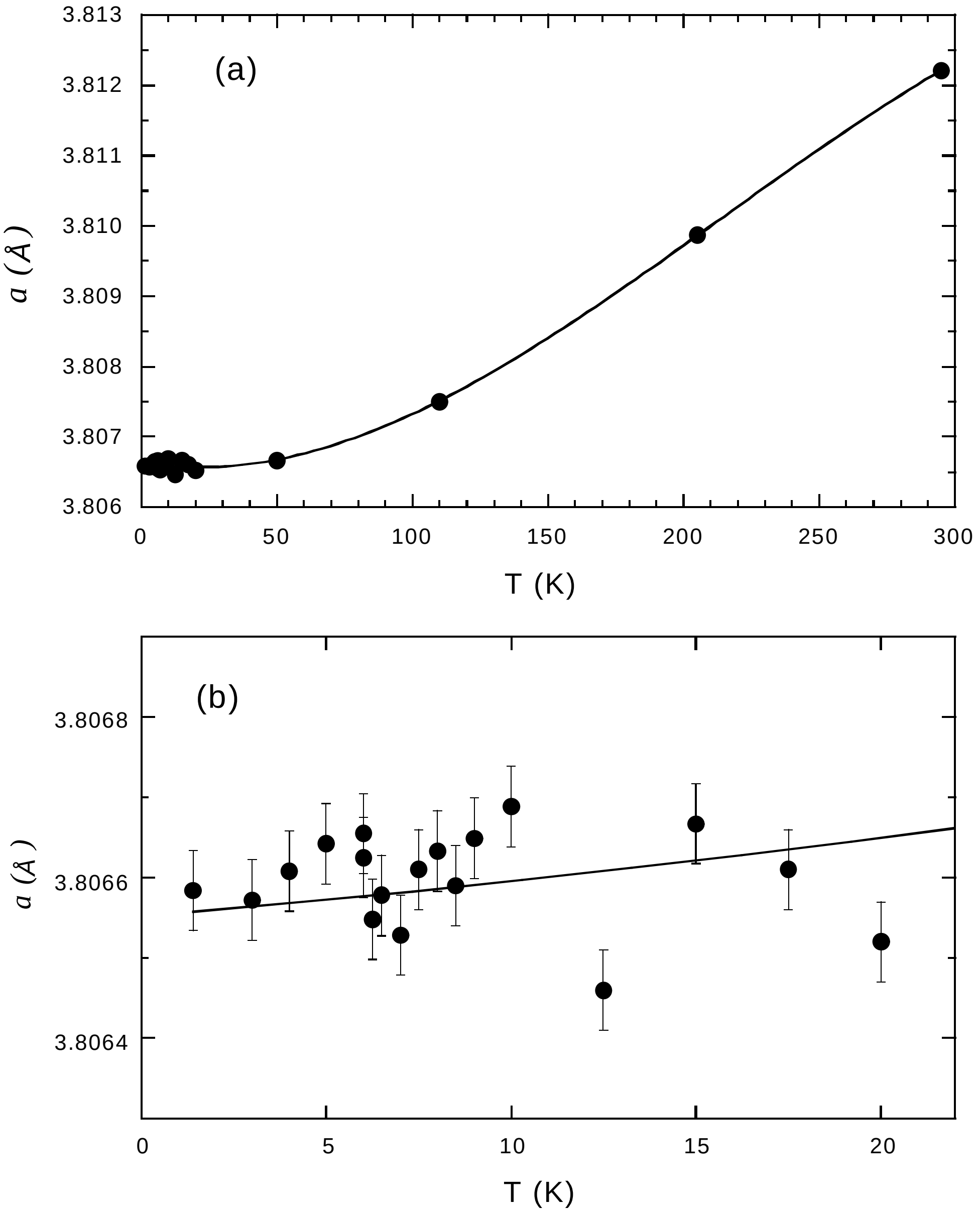

Figure 3 *Huang, et al*

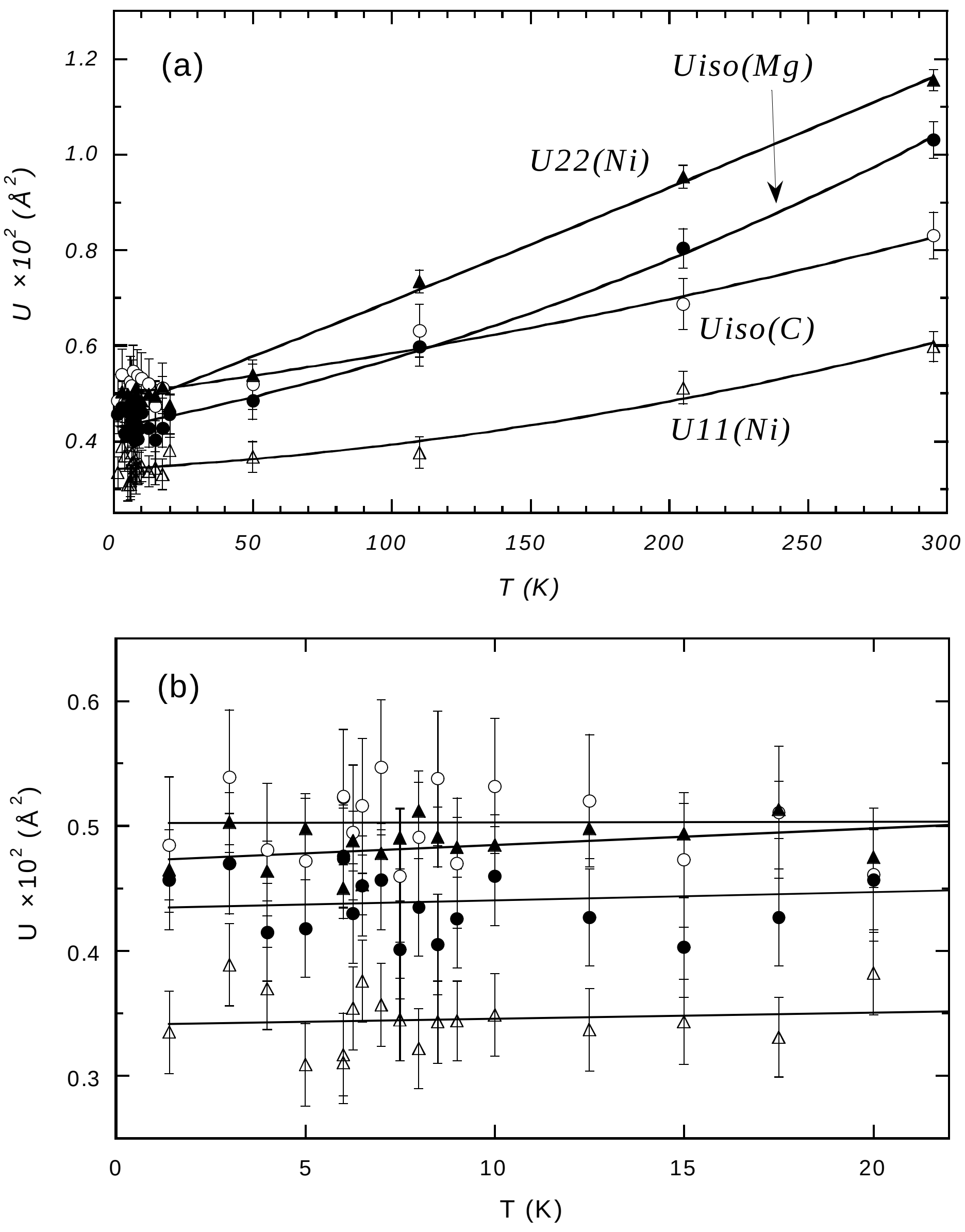

Figure 4 *Huang, et al*